\documentclass[aps, showpacs, showkeys,nofootinbib,floatfix]{revtex4}

\usepackage{amssymb}
\usepackage{amsmath}
\usepackage{graphicx}
\usepackage{hyperref}

\begin{document}

\title{Quantum Statistical Entropy and Minimal Length of $5D$ Ricci-flat Black String with Generalized Uncertainty Principle}

\author{Molin Liu}
\email{mlliudl@student.dlut.edu.cn}
\author{Yuanxing Gui}
\email{guiyx@dlut.edu.cn}
\author{Hongya Liu}
\email{hyliu@dlut.edu.cn}

\affiliation{School of Physics and Optoelectronic Technology, Dalian
University of Technology, Dalian, 116024, P. R. China}

\begin{abstract}
In this paper, we study the quantum statistical entropy in a $5D$
Ricci-flat black string solution, which contains a $4D$
Schwarzschild-de Sitter black hole on the brane, by using the
improved thin-layer method with the generalized uncertainty
principle. The entropy is the linear sum of the areas of the event
horizon and the cosmological horizon without any cut-off and any
constraint on the bulk's configuration rather than the usual
uncertainty principle. The system's density of state and free energy
are convergent in the neighborhood of horizon. The small-mass
approximation is determined by the asymptotic behavior of metric
function near horizons. Meanwhile, we obtain the minimal length of
the position $\Delta x$ which is restrained by the surface gravities
and the thickness of layer near horizons.
\end{abstract}


\maketitle


\section{Introduction}
One decade ago, several attempts on solving the gauge hierarchy problem,
based on higher dimensional gravity theories, were proposed (e.g. cite Arkani-Hamed-Dimopoulos-Davli model \cite{ADD} and Randall-Sundrum model \cite{RS1} \cite{RS2}).
According to the spirit of these brane world theory, we may live on a (3 + 1) dimensional hypersurface (3-brane) where the
standard model fields (such as fermions, gauge bosons, Higgs fields)
are confined without accessing along the transversal extra dimensions. The
branes are embedded in a higher dimensional space (bulk), in
which only gravitons and scalar particles without charges could
propagate. If the ordinary matter is
confined on the brane, the gravitational collapsing can not be
avoided. Hence a black hole can form naturally on the brane and
its horizon is extended along the extra dimensions. The collapsed
object which is a black hole on the brane is looked actually like
a string in the bulk \cite{Dblackhole}, if the total dimensionality is five. On the other hand, the
fact that the universe is accelerating expansion is justified by famous astronomical data of type
Ia Supernovae (SNe Ia) \cite{I} and the cosmic microwave background
(CMB) of the Wilkinson Microwave Anisotropy Probe (WMAP) \cite{cmb}.
Of course, there are many other key observations, including Two-degree-Field Galaxy
 Redshift Survey (2dFGRS) \cite{Colless} \cite{Peacock}, Sloan Digital
 Sky Survey (SDSS) \cite{Tegmark} and so on, also indicate the acceleration of the
universe. These observations in cosmology show that the space-time structure of our universe may be
 a de Sitter geometry in both the past and the future \cite{Witten}.
 In what follows, we consider a black string
solution which contains a $4D$ de Sitter geometry with an effective
cosmological constant offering an accelerating universe. This rich
class of $5D$ black brane solutions are found originally by Mashhoon
\cite{ref:Mashhoon} and restudied recently by many people focusing
mainly on the induced cosmological constant $\Lambda$
\cite{induconstant}, the extra dimensional force \cite{extraforce},
the radiative potential \cite{ref:Liu00}\cite{0607}, quasi-normal
modes \cite{QNMliu} and so on. In a recent work
\cite{entropyliu}, we used the two-dimensional area, which
leads to that we can not consider the contribution of extra
dimensional component in total quantum number, to describe the
black string's entropy with the standard uncertainty principle (UP).
The small-mass approximation is naturally obtained by the assumption
of far-apart two branes. In this paper, we will restudy this problem
based on a more general situation with the alternative generalized
uncertainty principle (GUP).

In 1970s, Bekenstein \cite{Bekenstain} and Hawking \cite{Hawking1}
showed that the black holes should be treated as thermodynamic
systems and the entropy of a black hole should be proportional to
the area of its horizon. After that, many literatures contributed to
the origin of black hole entropy by various approaches
\cite{Howtounderstand}. One famous approach is the brick-wall method
(BWM) shown by 't Hooft in 1985 \cite{Hooft}. In his work, 't Hooft
studied the quantum statistical property of scalar field in
Schwarzschild black hole and used the entropy of exterior field to
calculate the interior entropy. In fact, the equivalence of exterior
entropy and interior entropy comes from the explanation of entangled
entropy \cite{added}, i.e., the exterior field and interior field
are in the forms of pure state entanglement. So, the entropy of
black hole can be treated as the entropy of the contribution from
outside field. Certainly, the standard brick-wall method is used
only in the single horizon black hole space in which the external
field should be in thermal equilibrium. However, in the aspect of
the multi-horizon black hole space, the improved thin-layer method
(TLM) shown by Li et al in \cite{Li} is more applicable. Meanwhile,
the GUP, which is firstly applied to calculate entropy by Li
\cite{GUP} (the related works also can be found in \cite{relatedGUP}
\cite{RNGUP} \cite{Kimd}), is an effective method to solve
ultraviolet divergences in the vicinity of the horizon. In this
paper, we will use the method of GUP to restudy the quantum
statistical entropy in a $5D$ Ricci-flat black string space.

This paper is organized as follows: In Section 2, the $5D$
Ricci-flat black string metric and the surface gravities near
horizons are presented. In section 3, we calculate the proper total
number of quantum states including radial and extra dimensional
components under GUP. In section 4, the free energy and entropy of
this $5D$ Ricci-flat black string are calculated without
configuration assumption. Section 5 is the conclusion. We adopt the
signature $(-, +, +, +, +)$ and put $\hbar$, $c$ ,and $G$ equal to
unity. Greek indices $\mu, \nu, \ldots$ will be taken to run over
$0, 1, 2, 3$ as usual, while capital indices A, B, C, $\ldots$ run
over all five coordinates $(0, 1, 2, 3, 4)$.

\section{Scalar field in the $5D$ Ricci-flat black string coupled with effective $4D$ cosmological constant}
Before calculating the entropy, it is necessary to present the
background of the space-time. A static, three-dimensional spherically
symmetric line element with a $4D$ effective cosmological
constant takes the form \cite{ref:Mashhoon}
\cite{ref:Wesson}\cite{ref:Liu}
\begin{equation}
d s^{2}=\frac{\Lambda
\xi^2}{3}\left[-f(r)dt^{2}+\frac{1}{f(r)}dr^{2}+r^{2}\left(d\theta^2+\sin^2\theta d\phi^2\right)\right]+d\xi^{2}, \label{eq:5dmetric}%
\end{equation}
where $\xi$ is an open non-compact extra dimensional coordinate. The
metric function is $f(r) = 1 - 2M/r - \Lambda r^2/3$ and $M$ is the
black hole's mass on the brane. The part of this metric inside the
square bracket is exactly the same line-element as the $4D$
Schwarzschild-de Sitter (SdS) solution, which is bounded by two
horizons
--- an inner horizon (event horizon) and an outer horizon
(cosmological horizon). The parameter $\Lambda$ is an induced
cosmological constant which is reduced from the $5D$ to $4D$. The
metric (\ref{eq:5dmetric}) is Ricci-flat $R_{AB}=0$ and there is no
cosmological constant in $5D$ space. So one can actually treat this
$\Lambda$ as a parameter which comes from the fifth dimension.

In this Ricci-flat $5D$ brane world, the binary Randall-Sundrum type
branes system can be constructed via a coordinate transformation
$\xi=\sqrt{3 / \Lambda}\ exp(\sqrt{\Lambda / 3}\ y)$
\cite{ref:Liu00}. Hence, the metric (\ref{eq:5dmetric}) takes a
conformal form
\begin{equation}
d s^{2}=e^{2\sqrt{\frac{\Lambda}{3}}y}\left[-f(r)dt^{2}+\frac{1}{f(r)}dr^{2}+r^{2}\left(d\theta^2+\sin^2\theta d\phi^2\right)+dy^{2}\right],\label{eq:5dmetric-y}%
\end{equation}
where $0\leq y \leq y_1$ is the extra dimensional coordinate and
$y_1$ is essentially a compactification length of extra dimension.
It should be notice that the singularity $\xi = 0$ in original metric (\ref{eq:5dmetric}) is removed by this coordinate transformation.
Two branes locate at the endpoints of the fifth dimension, which
could be very small as in Randall-Sundrum 2-brane model \cite{RS1}
or very large as in Randall-Sundrum 1-brane model \cite{RS2} by
pushing the second brane to the infinity. On the fixed extra
dimensional hypersurface this metric describes a SdS black hole. But
it actually is a black string intersecting the brane world. The
horizon looks like a black string instead of a $4D$ sphere lying
along the fifth dimension. So we call the solution
(\ref{eq:5dmetric-y}) black string. The following calculation of
entropy will base on this new black brane solution (\ref{eq:5dmetric-y}).

In order to study the statistic thermodynamics feature in the
vicinity of horizons, it is necessary to show its space-time
structure near the horizons $r_e$ and $r_c$. The metric function
$f(r)$ in Eq. (\ref{eq:5dmetric-y}) can be recomposed as follows
\begin{equation}
f(r)=\frac{\Lambda}{3r}(r-r_{e})(r_{c}-r)(r-r_{o}). \label{re-f function}%
\end{equation}

The singularity of the metric (\ref{eq:5dmetric-y}) is determined by
$f(r)=0$. The real solutions to this equation are black hole event
horizon $r_{e}$, cosmological horizon $r_{c}$ and a negative
solution $r_{o}=-(r_{e}+r_{c})$. The last one has no physical
significance, and $r_{c}$ and $r_{e}$ are given as

\begin{equation}
\left\{
\begin{array}{c}
r_{c} = \frac{2}{\sqrt{\Lambda}}\cos\chi ,\\
r_{e} = \frac{2}{\sqrt{\Lambda}}\cos(\frac{2\pi}{3}-\chi),\\
\end{array}
\right.\label{re-rc}
\end{equation}
where $\chi=\frac{1}{3}\arccos(-3M\sqrt{\Lambda})$ with $\pi/6
\leq\chi\leq \pi/3$. The real physical solutions are accepted only
if
 $\Lambda$ satisfy $\Lambda M^2\leq\frac{1}{9}$ \cite{ref:Liu}.

The definition of the surface gravity is%
\begin{equation}
\kappa_{i}=\frac{1}{2}\left|\frac{df}{dr}\right|_{r=r_i}.
\end{equation}
These are
\begin{equation}
    \kappa_{e}=\frac{(r_{c}-r_{e})(r_{e}-r_{o})}{6r_{e}}\Lambda,\label{Ge}
\end{equation}
\begin{equation}
    \kappa_{c}=\frac{(r_{c}-r_{e})(r_{c}-r_{o})}{6r_{c}}\Lambda.\label{Gc}
\end{equation}
The different surface  gravities $\kappa_e$ and $\kappa_c$ give the
different temperatures $T_e = \frac{\kappa_e}{2\pi K_b}$ and $T_c =
\frac{\kappa_c}{2\pi K_b}$ near inner and outer horizons, where
$K_B$ is Boltzmann constant. It indicates that it is impossible
to use the original BWM to calculate the free energy in non-equilibrium thermodynamic system.
The improved TLM can solve this problem by considering a thin layer near the event horizon or
the cosmological horizon. The non-equilibrium large structure is replaced by the local thermodynamic states
where the thickness of layer is a very small quantity. This behavior is similar to the infinitesimal calculus to some degree.
Though the whole space is non-equilibrium, the region in the two layers can be generally treated as the
equilibrium state. So we can still use equilibrium state's approach
of the statistic physics. The TLM is an effective method to deal with
the multi-horizon space. Many literatures have addressed to the
works of Schwarzschild-de Sitter black hole \cite{SdS}, Kerr-de
Sitter black hole \cite{ker}, Vaidya black hole \cite{Li} and others
\cite{other}.

The minimally coupled quantum scalar field $\Phi$ with mass $m$ on
the background (\ref{eq:5dmetric-y}) satisfies
\begin{equation}
\frac{1}{\sqrt{g}}\frac{\partial}{\partial
x^{A}}\left(\sqrt{g}g^{AB}\frac{\partial}{\partial{x^{B}}}\right)\Phi - m^2\Phi = 0.\label{Klein-Gorden equation}%
\end{equation}
The modes of the scalar field can be decomposed as the separable
form $ \Phi = e^{-i\omega t} \Psi (r,\theta, \phi) L(y)$ and
$\omega$ is the particle energy. Then the equations for $L(y)$ and
$R_{\omega}(r)$ read as follows,

\begin{eqnarray}
       e^{-3\sqrt{\frac{\Lambda}{3}}y}\frac{d}{d y}\left (e^{3\sqrt{\frac{\Lambda}{3}}y}\frac{d}{d
    y}\right )L(y) + \left (e^{2\sqrt{\frac{\Lambda}{3}}y}m^2 + \mu^2\right
    )L(y) &=& 0,\label{L} \\
 \nonumber   \frac{\partial}{\partial r} \left(r^2 f (r) \frac{\partial}{\partial r}\right) \Psi + \frac{1}{\sin \theta} \frac{\partial }{\partial \theta} \left(\sin \theta\frac{\partial}{\partial \theta}\right)\Psi \ \ \ \ \ \ \ \ \ \ \ \ \ \ \ \ &&\\
    + \frac{1}{\sin\theta} \frac{\partial}{\partial \phi}\left(\frac{1}{\sin \theta}\frac{\partial}{\partial \phi}\right)\Psi +\left(\frac{\omega^2}{f(r)}-\mu^2\right)\Psi &=& 0,\label{r}
\end{eqnarray}
where the eigenvalue $\mu^2$ is the effective mass squared on the brane. In
previous work \cite{entropyliu}, the effective mass $\mu$ on the
brane is associated with the particle mass $m$ via RS
Arnowitt-Deser-Misner (ADM) mass relationship i.e. $\mu = m$ is on
the first brane and $\mu = m e^{\sqrt{\Lambda/3}y_1}$ is on the
second brane. In this paper, we do not consider the RS ADM mass
relationship but rather we will study this problem from the perspective of GUP.

\section{Proper total number of quantum states with energy less than $\omega$ under generalized uncertainty principle}
In order to calculate the total number of quantum states, we proceed
from two parts, one is the radial component $g_r(\omega)$ and the
other is the extra dimensional component $g_y(\omega)$, on the basis
of the corresponding Eqs. (\ref{L}) and (\ref{r}).

Substituting $\Psi \sim exp[i $S (r, $\theta$, $\phi$)$]$ into Eq.
(\ref{r}), we get the momentums's relationship by the effective WKB
approximation. Namely, the momentums $p_r = \frac{\partial
S}{\partial r}$, $p_{\theta} = \frac{\partial S}{\partial \theta}$
and $p_{\phi} = \frac{\partial S}{\partial \phi}$ satisfy
\begin{equation}\label{momentum}
    p_{r}^2 f(r) + \frac{p_{\theta}^2}{r^2} +
    \frac{p_{\phi}^2}{r^2\sin^2\theta} - \frac{\omega^2}{f(r)} +
    \mu^2 = 0.
\end{equation}
Comparing with the formal representations of other usual momentums
such as Schwarzschild black hole \cite{GUP}, Reissner-Nordstrom
black hole \cite{RNGUP}, Randall-Sundrum black string \cite{Kimd}
and others \cite{relatedGUP}, we find that they are very much alike
each other. The differences between this $5D$ Ricci-flat black
string and the others' are that the metric function $f(r)$ is
different and the parameter $\mu$ is the effective mass of $m$. More
importantly, there is a novel understand in the view of the higher
dimensional space.

The volume of momentum phase space is obtained as follows
\begin{eqnarray}\label{Vp}
   \nonumber V_p &=& \int d p_r d p_{\theta} p_{\phi} =\frac{4\pi}{3}\sqrt{\frac{1}{f(r)}p \cdot r^2 p \cdot r^2\sin^2\theta
    p}\\
     &=& \frac{4\pi r^2\sin\theta}{3\sqrt{f(r)}}\left(\frac{\omega^2}{f(r)}-\mu^2\right)^{3/2},
\end{eqnarray}
where $p$ is the module momentum which is determined by

\begin{equation}\label{module}
    p^2 = p_{i}p^{i}=g^{11}p_{r}^2 + g^{22}p_{\theta}^2 +
    g^{33}p_{\phi}^2 = \omega^2 f(r)^{-1} - \mu^2.
\end{equation}

In the standard quantum mechanics, the position $\hat{x}$ and the
momentum $\hat{p}$ are a pair of conjugate observable parameter
which satisfy the uncertainty principle
\begin{equation}\label{UP}
    \Delta x \Delta p \geq \frac{\hbar}{2}.
\end{equation}
It means that the uncertainty of position could be arbitrarily small
with the increasing uncertainty of momentum. However, in the quantum
system under the Planck scale the above UP relation should be
corrected to the generalized uncertainty
principle\cite{Garay}\cite{GUP}\cite{relatedGUP}\cite{RNGUP}\cite{Kimd}
\begin{equation}\label{GUP}
    \Delta x \Delta p \geq \frac{\hbar}{2}\left[1 + \gamma \left(\frac{\Delta
    p}{\hbar}\right)^2\right].
\end{equation}
This second-order equation gives a  minimal length
$\Delta x \geq \sqrt{\gamma}$ by the solution existence condition
$\Delta = \Delta x^2 - \gamma \geq 0$.

In the coordinate and momentum phase space, the phase space is
divided into a lot of ergospheres one by one, i.e., using
$\varepsilon \sim \varepsilon + d \varepsilon$ to describe the
quantum state. Furthermore, each ergosphere is also divided into smaller grids and each grid denotes one quantum state. According to the
GUP relation (\ref{GUP}), the grid or cell gives the length $2 \pi
\hbar (1 + \gamma p^2)$. Hence the number of quantum state in the
volume element $d^3 x d^3 p$ is
\begin{equation}\label{quan1}
\frac{d^3 x d^3 p}{(2\pi \hbar)^3(1+ \gamma p^2)^3}.
\end{equation}
 The corresponding quantum state density is
 \begin{equation}\label{quntumdensity}
    g (\varepsilon) = \frac{1}{(2\pi)^3}\int \frac{d r d \theta d\phi d p_r d p_{\theta} d p_{\phi}}{(1 + \gamma
    p^2)},
 \end{equation}
where we use the natural unit, i.e., $\hbar = c = 1$. Substituting
Eqs. (\ref{Vp}) and (\ref{module}) into the density
(\ref{quntumdensity}), we can get the number of quantum states
related to the radial modes as follows,
\begin{eqnarray}\label{quntumdensity2}
    \nonumber g_{r} (\varepsilon) &=& \frac{1}{(2\pi)^3} \int d r d \theta d \phi \left[1 + \gamma \left(\frac{\omega^2}{f(r)} -
    \mu^2\right)\right]^{-3} \int d p_r d p_{\theta} d p_{\phi}\\
    &=& \frac{2}{3\pi}\int dr \frac{r^2 \left[\omega^2/f(r) - \mu^2\right]^{3/2}}{f(r)^{1/2}\left[1 + \gamma \left(\omega^2 / f(r) - \mu^2\right)
    \right]^3}.
\end{eqnarray}
Here we analyse the asymptotic behavior of integrand near the event
horizon and cosmological horizon through the limit $f(r)
\longrightarrow 0 $. The integrand of Eq. (\ref{quntumdensity2}) is
clearly reduced to
\begin{equation}\label{integrand}
r^2\gamma^{-2} \omega^{-3} (1 - \frac{2M}{r} -
\frac{\Lambda}{3}r^2).
\end{equation}
Apparently, it is convergent near the two horizons. This is ensured
by GUP and the similar behavior can be found in many works
\cite{GUP} \cite{relatedGUP} \cite{RNGUP} \cite{Kimd}.

In the previous work \cite{entropyliu}, two-dimensional area is used
to describe this black string's entropy and the
contribution of extra dimensional quantum number was not considered.
Then, in this paper we employ the total quantum number, which is
composed by the radial part and the extra dimensional part, to
calculate the system's free energy $F(\beta)$.

The radial part $g_{r}(\omega)$ is obtained in the above expression
(\ref{quntumdensity2}) and the extra dimensional $g_{y}(\omega)$ can
be derived directly from the corresponding Eq. (\ref{L}). Through
the WKB approximation we substitute $L (y) \sim exp[i$Y(y)$]$ into
Eq. (\ref{L}) and keep the real part. So the wave number $k_y =
\frac{\partial Y}{\partial y}$ satisfies
\begin{equation}\label{ky}
k_y^2 = e^{2\sqrt{\frac{\Lambda}{3}}y}m^2 + \mu^2.
\end{equation}
The quantum number of the fifth dimensional mode is given by
\begin{eqnarray}
  \nonumber g_{y} (\omega) &=& \frac{1}{\pi} \int_{m}^{\omega/\sqrt{f}} d \mu \int_{0}^{y_1} dy \frac{\partial k_y (y,\mu)}{\partial \mu}\\
   &=& \frac{1}{\pi} \sqrt{\frac{3}{\Lambda}} \int_{m}^{\omega/\sqrt{f}} d \mu \left(ArcCoth \frac{\mu}{\sqrt{1 + \mu^2}} - ArcCoth \frac{\mu}{\sqrt{e^{2y_1\sqrt{\Lambda/3}} + \mu^2}}\right).\label{yquannumb}
\end{eqnarray}

Accordingly, the proper total quantum number $g_{t}(\omega)$ with
energy less than $\varepsilon$ can be given formally as follows,
\begin{equation}\label{total}
    g_{t}(\omega) = \int d g_{t} (\omega) = \int d g_r (\omega) d g_y (\omega).
\end{equation}

\section{Convergent free energy and entropy of black string without cut-off}

According to the bosons' ensemble theory, the free energy in terms of the inverse temperature $\beta$ is written as
\begin{eqnarray}\label{freeenergy1}
    \nonumber F &=& \frac{1}{\beta} \sum _{n} \ln \left[1 - e^{-\beta
    \omega_{n}}\right] \approx \frac{1}{\beta} \int d g_{t} (\omega) \ln (1 - e^{-\beta \omega})\\
\nonumber &=& -\int_{\mu \sqrt{f}}^{+\infty} d \omega
\frac{g_{t}(\omega)}{e^{\beta \omega} - 1}\\
&=& - \frac{2}{\sqrt{3\Lambda}\pi^2}\int_{\mu\sqrt{f}}^{\infty} d
\omega \frac{1}{e^{\beta \omega} - 1} \int _{r} dr
\frac{r^2}{\sqrt{f(r)}} \int_{m}^{\omega/\sqrt{f}} d \mu \ \xi
(\mu),
\end{eqnarray}
where
\begin{equation}\label{xi}
    \xi (\mu) = \frac{\left(\frac{\omega^2}{f(r)} - \mu^2\right)^{3/2}}{\left[1 +
\gamma \left(\omega^2 / f(r) - \mu^2\right)\right]^3}\left[ArcCoth
\frac{\mu}{\sqrt{1 + \mu^2}} - ArcCoth \frac{\mu}{\sqrt{e^{2 y_1
\sqrt{\Lambda/3}} + \mu^2}}\right].
\end{equation}
The summation can be rewritten as an integral form with the
continuum limit. The compressed $\gamma$ term, which is derived
directly by GUP, has been taken to ensure the convergence of free
energy near horizons. Meanwhile, there is a limit of
$f(r)\rightarrow 0$ in the vicinity of the horizon. So the term
$\omega^2 / f(r) - \mu^2$ can be transferred to $\omega^2 / f(r)$.
Hence, the function $\xi (\mu)$ can be rewritten as a new expression
near both horizons
\begin{equation}\label{xi1}
    \xi (\mu) = \frac{\omega^3}{f(r)^{3/2} \left(1 + \gamma \frac{\omega^2}{f(r)}\right)^3}\left[ArcCoth
\frac{\mu}{\sqrt{1 + \mu^2}} - ArcCoth \frac{\mu}{\sqrt{e^{2 y_1
\sqrt{\Lambda/3}} + \mu^2}}\right].
\end{equation}
 This approximation simplifies greatly the calculation of the integration about variable $\mu$. The first
fraction can be treated as a constant to deal with the $\mu$ aspect.
Hence, the last integration of free energy (\ref{freeenergy1}) is
given as
\begin{eqnarray}\label{xi2}
    \nonumber \int_{m}^{\omega/\sqrt{f}} d \mu \xi(\mu) &=& \frac{\omega^3}{f(r)^{3/2} \left(1 + \gamma
    \frac{\omega^2}{f(r)}\right)^3} \bigg{\{}\sqrt{e^{2 y_1 \sqrt{\frac{\Lambda}{3}}} + \mu^2} -\sqrt{1 + \mu^2} \\
   \nonumber  &+& \mu ArcCoth \frac{\mu}{\sqrt{1 + \mu^2}} - \mu ArcCoth \frac{\mu}{\sqrt{e^{2y_1 \sqrt{\frac{\Lambda}{3}}} +
     \mu^2}}\bigg{\}} \bigg{|}_{m}^{\omega/\sqrt{f}}\\
     &=&\frac{\omega^3 \alpha}{f(r)^{3/2} \left(1 + \gamma
    \frac{\omega^2}{f(r)}\right)^3},
\end{eqnarray}
where
\begin{equation}\label{alpha}
    \alpha =  \sqrt{1 + m^2} - \sqrt{e^{2y_1\sqrt{\frac{\Lambda}{3}}} + m^2} + m \cdot ArcCoth \frac{m}{\sqrt{e^{2y_1\sqrt{\frac{\Lambda}{3}}}+m^2}} - m\cdot ArcCoth \frac{m}{\sqrt{1 +
    m^2}}.
\end{equation}
It should be noticed that in the last step we use the limit $f(r)
\longrightarrow 0$. Substituting integral result (\ref{xi2}) to Eq.
(\ref{freeenergy1}) and transferring energy range $[\mu\sqrt{f},
+\infty]$ to $[0, +\infty]$, we obtain the free energy of this
system
\begin{equation}\label{freeenergy}
    F = -\frac{2\alpha}{\sqrt{3\Lambda}\pi^2}\int_{r}d r
    \frac{r^2}{f^2(r)} \int_{0}^{+\infty} d \omega \frac{\omega^3}{(e^{\beta \omega}-1)\left(1 + \gamma \omega^2 /
    f(r)\right)^3}.
\end{equation}
Hence, the scalar field's entropy in this $5D$ Ricci-flat black
string can be written directly from the derivative of free energy
with respect to $\beta$
\begin{eqnarray}\label{entropy}
   \nonumber S &=& \beta^2\frac{\partial F}{\partial \beta} = \frac{2\alpha}{\sqrt{3\Lambda}\pi^2} \int_{r} d r
\frac{r^2}{f^2(r)} \int_{0}^{+\infty}\frac{\omega^4}{\left(1 +
\gamma \frac{\omega^2}{f(r)}\right)^3}\cdot \frac{\beta^2 e^{\beta
\omega}}{(e^{\beta\omega} - 1)^2}\\
&=& \frac{2\alpha}{\sqrt{3\Lambda}\pi^2\gamma^{3/2}}\int_{r} d r
\frac{r^2}{f^{1/2}(r)} \int_{0}^{+\infty} d\zeta \frac{a^2
\zeta^4}{\left(e^{a \zeta / 2} - e^{- a \zeta / 2}\right)^2 (1 +
\zeta^2)^3},
\end{eqnarray}
where we adopt the Kim's coordinate transformation \cite{Kimd}
\begin{eqnarray}
  \zeta &=& \omega \sqrt{\frac{\gamma}{f}}, \label{zet}\\
  a &=& \beta \sqrt{\frac{f}{\gamma}}. \label{a}
\end{eqnarray}
This transformation will give the deterministic expression about the
last integral of variable $\omega$ in the expression of entropy
(\ref{entropy}). In the improved brick-wall method, the quantum
field is restricted in the vicinity of the horizons. The thin-layer
BWM boundary conditions are
\begin{eqnarray}
  \nonumber \Phi (t,\ r,\ \theta,\ \phi,\ y) &=& 0\ \ \ \ for \ \  r_e \leqslant r \leqslant r_e + \varepsilon_e,\\
  \Phi (t,\ r,\ \theta,\ \phi,\ y) &=& 0\ \ \ \ for \ \  r_c -
  \varepsilon_c\leqslant r \leqslant r_c,\label{boundarycondition}
\end{eqnarray}
where the thickness of layers is a small quantity, i.e., $r_e \gg
\varepsilon_e$ and $r_c \gg \varepsilon_c$. Hence, according to the
expression (\ref{a}) and the boundary conditions
(\ref{boundarycondition}), we can get that $a$ goes to zero for $r
\longrightarrow r_e$ or $r \longrightarrow r_c$. The integral about
variable $\zeta$ in Eq. (\ref{entropy}) is reduced to
\begin{equation}\label{zeta11}
    \int_{0}^{+\infty} d\zeta \frac{a^2
\zeta^4}{\left(e^{a \zeta / 2} - e^{- a \zeta / 2}\right)^2 (1 +
\zeta^2)^3} = \int _{0}^{ + \infty} d \zeta \frac{\zeta^2}{(1 +
\zeta^2)^3} = \frac{\pi}{16}.
\end{equation}

Finally, from the metric (\ref{eq:5dmetric-y}) the minimal length is
obtained as
\begin{eqnarray}\label{Minilength}
    \nonumber \Delta x &\geq& \sqrt{\gamma} = \int_{r_e}^{r_e + \varepsilon_e} +
    \int_{r_c - \varepsilon_c}^{r_c} \frac{d r}{\sqrt{f}}\\
\nonumber &=& \sqrt{\frac{3 r_e}{\Lambda (r_c - r_e) (r_e - r_o)}}
\int_{r_e}^{r_e + \varepsilon_e} \frac{d r}{\sqrt{r - r_e}} +
\sqrt{\frac{3 r_c}{\Lambda (r_c - r_e) (r_c - r_o)}} \int_{r_c -
\varepsilon_c}^{r_c} \frac{d r}{\sqrt{r_c - r}}\\
&=& \sqrt{\frac{2\varepsilon_e}{\kappa_e}} +
\sqrt{\frac{2\varepsilon_c}{\kappa_c}},
\end{eqnarray}
where $\kappa_e$ and $\kappa_c$ are the surface gravities on the
horizons $r_e$ and $r_c$, which are defined by expressions
(\ref{Ge}) and (\ref{Gc}) respectively. It is clear that the
contributions of the minimal length come from the sum of inner
horizon and outer horizon.

Substituting Eqs. (\ref{zeta11}) and (\ref{Minilength}) into Eq.
(\ref{entropy}), we obtain the final $5D$ black string's entropy
\begin{eqnarray}\label{entropy11}
\nonumber S &=& \frac{\alpha}{8\sqrt{3\Lambda}\pi\gamma^{3/2}}
\int_{r_e}^{r_e + \varepsilon_e} +
    \int_{r_c - \varepsilon_c}^{r_c} d r \frac{r^2}{\sqrt{f(r)}}\\
\nonumber &=& \frac{\alpha}{8\sqrt{3\Lambda}\pi\gamma^{3/2}} \cdot
\left(r_e^2 \sqrt{\frac{2\varepsilon_e}{\kappa_e}} + r_c^2
\sqrt{\frac{2\varepsilon_c}{\kappa_c}}\right)\\
&=& \frac{\alpha}{16\sqrt{6\Lambda}\pi^2
\gamma^{3/2}}\left(\sqrt{\frac{\varepsilon_e}{\kappa_e}}A_e +
\sqrt{\frac{\varepsilon_c}{\kappa_c}}A_c\right),
\end{eqnarray}
where $A_e = 4\pi r_e^2$ and $A_c = 4\pi r_c^2$ are the areas of
black hole horizon and cosmological horizon respectively. Comparing
with the previous result \cite{entropyliu}, this one shows more
clearly that the entropy is the linear sum of the area of the event
horizon and the cosmological horizon without cut-off factors. It
should be noticed that we do not give any constraint on the branes
including tension, configuration, ADM mass and so on. This result
also justifies that the solution on the brane is indeed a SdS black
hole when the $5D$ Ricci-flat black string with a $4D$
cosmological constant is reduced to the $4D$ space-time. The quantum
statistical entropy not only varies directly with total areas but
also is related to the thickness of the bulk and the field form.

\section{Conclusion}
We have restudied statistical mechanical entropies of a $5D$
Ricci-flat black string, which contains a $4D$ Schwarzschild-de
Sitter black hole on the brane, by using improved thin-layer BWM
under GUP. We obtain the convergent free energy and a minimal length
without any cut-off near two horizons which are unsolvable in the
classical uncertainty principle scenario.

In usual quantum mechanics, the position-momentum uncertainty
relation i.e. Hersenberg uncertainty does not consider the gravity. However, if the gravitational effect is
considered, there is an unavoidable minimal length being
proportional to Planck scale $L_{p}$ in the string theory. As the generalization of uncertainty relation, the GUP tell us some new aspects of quantum systems. Many efforts \cite{Kempf} have been devoted
to GUP and its consequences. It is known that GUP can avoid the
divergence of quantum number and free energy in the external quantum
field outside black hole, so the unnatural cut-off factor in UP is
removed in this paper.

In fact, the $5D$ Ricci-flat black string is different from the
usual higher dimensional space. There are two different temperature
horizons in this space, one is the inner event horizon and the other
is the outer cosmological horizon. It evidently shows that this
black string is a nonequilibrium system as a whole. Since the usual
BWM is only suitable for the thermal equilibrium state, it can
not be employed to study the entropy here. However, it is well known that
the thin-layer method can overcome this difficulty with GUP. The
related domain, which can be looked as an equilibrium state
completely, is only a thin layer in Planck scale near horizons.

The last result (\ref{entropy11}) includes a key parameter $\alpha$
coupled with the magnitude of extra dimension $y_1$, the
cosmological constant $\Lambda$ and the mass of particle $m$. If
$\alpha$ is negative, a singular negative entropy will appear.
However, though there is a relationship
\begin{equation}\label{alpha-11}
\sqrt{1 + m^2} \leqslant \sqrt{e^{2 y_1 \sqrt{\frac{\Lambda}{3}}} +
m^2}
\end{equation}
 in Eq.(\ref{alpha}).
But we know that the ArcCoth function is a monotone decreasing one
in the positive independent variable range and the last two terms
satisfy
\begin{equation}\label{alpha-22}
 m \cdot ArcCoth
\frac{m}{\sqrt{e^{2y_1\sqrt{\frac{\Lambda}{3}}}+m^2}} \geqslant
m\cdot ArcCoth \frac{m}{\sqrt{1 + m^2}}.
\end{equation}
So $\alpha$ could obtain a positive value theoretically. Because
there are four parameters i.e. $\alpha$, $\Lambda$, $m$ and $\mu$ in
the Eq.(\ref{alpha}), it is very difficult to solve distinctly the
inequality $\alpha > 0$ from four free variables. In fact, we can
look inequality $\alpha > 0$ as a constraint condition of black
string.

The GUP parameter $\gamma$ associated with string theory scale ---
Planck scale $l_{p}$ plays a key role in this model. As we all know,
the original GUP is used to explain the fact that string cannot
probe distances below the minimum length (its value is
$\sqrt{\gamma}$ in this paper). In the recent years, many people,
such as Li \cite{GUP}, Kim \cite{Kimd} and the others
\cite{relatedGUP} \cite{RNGUP}, use it to calculate black hole's
entropy and find this can obviate the cut-off needed in the usual
brick wall model. As we see in this paper, the volume element of
phase space changes through the GUP parameter $\gamma$. After
integral in the whole momentum space, referring to equation
(\ref{quntumdensity2}), we find the free energy is convergence in
the vicinity of the horizon since there is a suppressing
$\gamma$-term in the denominator induced from the generalized
uncertainty principle. So this convergence free energy ensures the
convergence entropy without the artificial cut-off. Else, comparing
the early work \cite{entropyliu} with this GUP model, the former
just is a particular limit that the second brane is push to faraway
places. Furthermore, the result (\ref{entropy11}) contains the
string theory scale, which also agree with the strong coupling $E_8
\times E_8$ heterotic string theory. Partly, this result supports
the statement that when calculating the density of quantum states we
should take into account the contribution from the string excitation
\cite{stringexcited}.

\acknowledgments Project supported by the National Basic Research
Program of China (2003CB716300), National Natural Science Foundation
of China (10573003) and National Natural Science Foundation of China
(10573004).

\end{document}